\documentclass[10pt]{aastex63}
\usepackage[utf8]{inputenc}
\usepackage{amsmath, float, graphicx}
\usepackage[normalem]{ulem}

\begin{document}

\title{A Spherical Shells Model of Atmospheric Absorption for Instrument Calibration}

\author{Nicolas Donders}
\affil{The University of Alabama in Huntsville, Huntsville AL}
\affil{The Department of Space Science and The Center for Space Plasma and Aeronomic Research, Huntsville AL}

\author{Genevieve Vigil}
\affil{NASA Marshall Space Flight Center, Huntsville AL}

\author{Adam Kobelski}
\affil{NASA Marshall Space Flight Center, Huntsville AL}

\author{Amy Winebarger}
\affil{NASA Marshall Space Flight Center, Huntsville AL}

\author{Larry Paxton}
\affil{Johns Hopkins Applied Physics Laboratory, Laurel MD}

\author{Charles Kankelborg}
\affil{Montana State University, Bozeman MT}

\author{Gary Zank}
\affil{The University of Alabama in Huntsville, Huntsville AL}
\affil{The Department of Space Science and The Center for Space Plasma and Aeronomic Research, Huntsville AL}

\begin{abstract}

We present a model for atmospheric absorption of solar ultraviolet (UV) radiation. 
The initial motivation for this work is to predict this effect and correct it in Sounding Rocket (SR) experiments.
In particular, the Full-sun Ultraviolet Rocket Spectrograph (FURST) is anticipated to launch in mid-2023.
FURST has the potential to observe UV absorption while imaging solar spectra between 120-181 nm, at a resolution of $\mathcal{R} > 2 \cdot 10^{4}$ ($\Delta V < \pm 15$ km/s), and at altitudes of between $\approx$ 110-255 km.
This model uses estimates for density and temperature, as well as laboratory measurements of the absorption cross-section, to predict the UV absorption of solar radiation at high altitudes.
Refraction correction is discussed and partially implemented but is negligible for the results presented.
Absorption by molecular Oxygen is the primary driver within the UV spectral range of our interest.
The model is built with a wide range of applications in mind.
The primary result is a method for inversion of the absorption cross-section from images obtained during an instrument flight, even if atmospheric observations were not initially intended.
The potential to obtain measurements of atmospheric properties is an exciting prospect, especially since sounding rockets are the only method currently available for probing this altitude \textit{in-situ}.
Simulation of noisy spectral images along the FURST flight profile is performed using data from the High-Resolution Telescope and Spectrograph (HRTS) SR and the FISM2 model for comparison.
Analysis of these simulated signals allows us to capture the Signal-to-Noise Ratio (SNR) of FURST and the capability to measure atmospheric absorption properties as a function of altitude.
Based on the prevalence of distinct spectral features, our calculations demonstrate that atmospheric absorption may be used to perform wavelength calibration from in-flight data.

\end{abstract}
\section{Introduction}\label{sec:intro}

Atmospheric absorption, in particular from Oxygen and Nitrogen molecules, significantly limits our ability to observe the Ultraviolet (UV) spectrum from the ground \citep{friedman1951}.
Many teams use sounding rockets, weather balloons, or high-altitude observatories for aeronomy or solar physics studies to get above most of the atmosphere.
Correcting for this absorption is possible, however, some light will always be lost.

There are many models that predict the properties of Earth's upper atmosphere.
The Naval Research Laboratory Mass Spectrometer and Incoherent Scatter Exosphere 2000 (NRLMSISE-00) atmospheric model is typically used to obtain the density of elements and their temperatures as they vary with altitude \citep{NRLMSISE}.
The most common absorption coefficient database, The High-Resolution Transmission (HITRAN) molecular spectroscopic database is a common resource for obtaining absorption coefficients, but it does not contain absorption cross-section information for O$_2$ in the UV spectral range \citep{gordon2017hitran2016}.
The Max-Planck Institute for Chemistry in Mainz, Germany (MPI-Mainz) Ultraviolet-Visible (UV-VIS) database \citep{MPI_Mainz_Database, keller2013mpi, sander2014mpi} does contain this data, however, it requires significant post-processing.

In our research, there are very few complete models for UV absorption and even fewer that cover wavelengths below 300 nm \citep{smette2015molecfit, kausch2015molecfit, jones2013advanced, noll2012atmospheric, lean1980atmospheric}.
To give an example, \citet{he2019radiative, sun2018reevaluating, sun2022optimal} aims to calculate the high-altitude atmospheric emission effects due to ``day-glow.''
However, it does this by looking through Earth's upper atmospheric layers.
This absorption model was designed for the specific purpose of determining the optical depth between two orbiting spacecraft and inferring the abundance of certain molecules in that level of the upper atmosphere.
Many aeronomy models account for the self-absorption and re-emission of light, and more information can be found in \cite{degenstein2000atmospheric}.

A pressing issue with all of these available models is that they are closed-source and platform-specific programs.
It would be extremely beneficial to the atmospheric community if a general-purpose program could be developed that pulled together these various resources.
Thus, we have developed an open-source platform-independent model with the primary goal of being more broadly applicable.
This goal is in line with other National Science Foundation (NSF) goals such as the EarthCube project led by the National Center for Atmospheric Research (NCAR) \citep{abernathey2017EarthCube}.

As a reference for building our model, a review paper by Meier (1991) discusses many models of the time and presents a plot of the ``Altitude of Unit Optical Depth'' (\cite{meier1991ultraviolet}, Figure 2).
Our initial aim was to match these results by combining the NRL density model and the MPI-Mainz UV-VIS database, providing a platform that can be adapted to fit the user's needs.

The initial inspiration for this work was to understand atmospheric absorption during sounding rocket flights, for example, the upcoming Full-sun Ultraviolet Rocket Spectrograph (FURST). FURST was motivated by a distinct lack of complete Vacuum Ultraviolet (VUV, 120-181 nm) spectra at high-resolution ($\mathcal{R}>2\cdot10^{4}$, $\Delta V < 15$ km/s) of the full solar disk.
A good example of the currently available data is SUMER, which has a good resolution ($\mathcal{R} \approx 7-16 \cdot 10^{4}$ from 67-161 nm) but did not capture the full disk (1'' x 300'' slit) \citep{curdt2001sumer, peter1999analysis, peter1999doppler}.
SUMER can perform a raster across the surface, but it takes 31 hours to do so, limiting the temporal resolution of such a measurement.
This and other slit-based spectrometers focus on narrow wavelength bands and cannot capture the full disk of the Sun over small time scales.
FURST will provide not only an opportunity to observe the full disk of the Sun in such an important and wide wavelength range but should observe UV absorption from Earth's upper atmosphere.

In addition, we intend to launch a set of instruments from higher latitudes.
The Hi-C Flare campaign will require waiting on the launch pad for hours in order to launch at the opportune time to capture an active flaring region on the Sun \citep{savage2021hicflare}.
With this constraint, Poker Flat, Alaska is required to be used.
However, at this latitude and extended launch windows, some atmospheric absorption will be inevitable.
Predicting this effect will assist our team in adjusting the launch windows (time of year/day) for this exciting campaign.

In this paper, we discuss the development of an atmospheric absorption model that can be used for various purposes.
To test one application, we simulate the potential impact on an instrument and utilize specific absorption features as fiducial points to aid in spectral calibration.
Additionally, given a high spectral and radiometric resolution, we have the capability to analyze spectral images of past sounding rockets to uncover the fundamental physical properties of absorption-causing elements.
\textit{In-situ} measurements of the kind and at these altitudes can only be done with sounding rockets.

In Section \ref{sec:methods}, we introduce the methods for building up the spherical shells model, including refraction and atmospheric properties.
We set up the model by introducing the optical depth calculation for a curved atmosphere with refraction.
The model is then simplified for computation with imported databases for atmospheric density and molecular absorption cross-section.
We finish the model by simulating a spectral image on our sounding rocket instrument with these absorption effects.
Finally, we discuss two possible results from such an approach: spectral calibration and inversion of atmospheric information from raw data.
In Section \ref{sec:results} we will summarize these results, and in Section \ref{sec:discussion} we discuss the potential uses for this technique and future work.
\section{Methods}\label{sec:methods}

\subsection{Differential Path Length}\label{sec:diffpath}

To introduce the necessary components for building up our model, we start with the optical depth.
This term characterizes the amount of material that light will have to pass through before arriving at the detector.
One normally uses Beer-Lambert's Law to calculate the optical depth $\tau$, which requires knowledge of the absorption cross-section $\sigma$ (cm\textsuperscript{2}/molecule) and the number density $\eta$ (molecules/cm\textsuperscript{3}) as follows:

\begin{align}\label{eqn:opticaldepth}
    I
    = I_{0} e^{-\tau}, \
    \tau (r, \lambda)
    = \int_{r}^\infty \sigma \left( r, \lambda \right) \eta \left( r' \right) \frac{ dr' }{ \mu }
    \implies \tau_{i,j}
    = \sum_{i'=i}^{\infty} \sigma_{i',j} \eta_{i'} \Delta L_{i},
\end{align}
where $I$ is the intensity at the detector, $I_{0}$ is the intensity of the source.
The optical depth varies with height $r$ (km) and wavelength $\lambda$ (nm).
Since we are developing a database-driven model, we switch to index notation and label them as $i$ and $j$, respectively.
We write the transmission factor $T=I/I_0$ to measure how much light would be absorbed in the specified altitudes and wavelengths.

\begin{figure}
    \centering
    \includegraphics[width=0.9\linewidth]{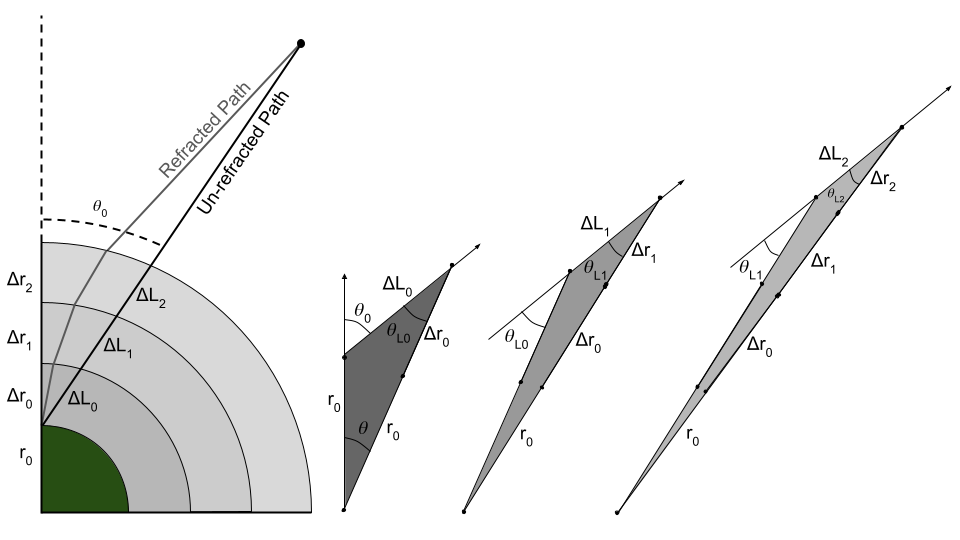}
    \caption{Cartoon of Geometry and Trigonometry to solve (not to scale). The spherical shells have an initial zenith angle that depends upon the starting point of the observation, $r_0$. The differential path length $\Delta L_i$ is labeled for a refraction/refraction-less setup. The next 3 diagrams show the relevant geometries at this initial height, with the effect of refraction exaggerated.
    The first triangle shows the necessary angles and side lengths in order to solve for $\Delta L_0$.
    The next two show the progression in solving for $\Delta L_i$ along each additional shell.
    For the proper optical depth calculation in Equation \ref{eqn:opticaldepth}, we must calculate all intermediate $\Delta L_i$ along the optical path at each subsequent height $r_i$.
    }
    \label{fig:diagram_trig}
\end{figure}

It is the convention to denote the zenith angle $\theta$ in the form $\mu \equiv \cos \theta$.
In traditional models, the factor $\mu$ alone can account for the change to optical depth due to changing zenith angle (see left diagram in Figure \ref{fig:diagram_trig}).
This is valid for a small angle $\theta$, however, we want a more robust model that takes into account the curvature of the Earth's atmosphere, refraction, and changing altitude.
With this in mind, we have combined it with the model altitude resolution $\Delta r$ and will refer to it as the ``differential path length,'' $\Delta L_{i} \equiv \Delta r_{i} / \mu_{i}$.

Solving for the differential path length is a primary contribution of this model.
The diagram for setting up the geometrical argument is shown in Figure \ref{fig:diagram_trig}.
To add further complexity, we have considered the cross-section $\sigma$ to be a function of temperature.
This dependence becomes very important at the altitude of unit optical depth, as our later results will highlight.
Temperature is of course dependent upon altitude, so we write $\sigma \equiv \sigma \left( r, \lambda \right)$.
We also attempt to account for the impact of refraction to ensure we have covered the most significant factors.

Using these given variables, we can calculate the adjusted $\Delta L$ as described in Figure \ref{fig:diagram_trig}.
In this Figure, the spherical shells are shown on the left, along with 3 triangles that highlight the relevant trigonometry.
Beginning at ground level, we solve for each $\Delta L$ along the path length from the observer to the source.
The triangles here indicate the necessary angles and side lengths needed to solve for each $\Delta L_i$.
As in Equation \ref{eqn:opticaldepth}, these non-uniform differential path lengths are what distinguish this setup from the simple model with uniform $\Delta L = \Delta r / \mu$.

There are many ways to solve these triangles for $\Delta L_i$, but we have done so in a way that requires the intermediate angle $\theta_{Li}$ in order to prepare for the application of refraction.
To solve for the differential path length $\Delta L_{i}$, we employ simple trigonometric relations (see \ref{sec:appendixdiffpathlength}) to work out that $\Delta L_0 = \sqrt{ \left( r_0 + \Delta r_0 \right)^{2} - r_{0}^{2} \left( 1 - \mu_0^2 \right) } - r_0 \mu_0$.
This is the solution for the first spherical shell at the initial altitude $r_{0}$.
Extending to the generalized solution for each subsequent shell $i$ in the differential path length $\Delta L_{i}$:

\begin{align}
    \Delta L_i
    &= \sqrt{ \left( r_0 + \sum_{i'=0}^i \Delta r_{i'} \right)^{2} - \left( r_{0} + \sum_{i'=0}^{i-1} \Delta r_{i'} \right)^{2} \left( 1 - \mu_i^2 \right) } - \left( r_{0} + \sum_{i'=0}^{i-1} \Delta r_{i'} \right) \mu_i, \\
    \mu^{2}_{i}
    &= 1 - \left( \frac{ r_0 + \sum_{i'=0}^{i-1} \Delta r_{i'} }{ r_0 + \sum_{i'=0}^{i} \Delta r_{i'} } \right)^2 \left( 1 - \mu_{i-1}^2 \right), \label{eqn:ThetaLi}
\end{align}
giving us a solution for the length of each discrete path between all shells.
This solution is fed into Equation \ref{eqn:opticaldepth} to yield the total absorption at each particular wavelength for the starting altitude $r_{0}$.
We can then adjust the starting altitude $r_{0}$ and perform the calculations again to build up the model for the absorption profile.

As mentioned in Section \ref{sec:intro}, a popular paper that describes atmospheric absorption in these wavelengths is \cite{meier1991ultraviolet}.
We set the date, time, and location listed in Table 6 from that paper: March 21, 1980, 10:00 AM local time, at the GPS coordinates for White Sands Missile Range (WSMR), NM.
From those details, we can query various databases and online resources for the initial zenith angle $\mu_0 \approx 0.7$ (from PySolar \cite{brandon_stafford_2018_1461066}), and the altitude and sea level for $r_0 \approx 6379.4$ km (from \cite{USGS2021}).
We also load into the program the PyNRLMSISE-00 atmospheric model data at a defined resolution $\Delta r = 1$ km \citep{Bender2020PyNRLMSISE00}.
Higher spatial resolutions can of course be specified as desired, but show no appreciable difference to the results presented in this paper.

In the calculations above, $\theta_{Li}$ is solved at each step and depends on the normal angle from the previous shell.
This was done intentionally to provide a natural progression to refraction correction.
\subsection{Refraction}\label{sec:refraction}

We refer to \cite{born2013principles} for a rigorous derivation, showing that refraction corrections are small for altitudes above 20 km and do not exceed 0.2$^{\circ}$.
For completeness, we have developed the tools for applying refraction to the intermediate angle $\theta_{Li}$ at the shell boundaries.
For the applications in this paper, it is not necessary to apply these corrections
Refer to \ref{sec:appendixdiffpathlength} for more details.

From Snell's law, there exists an equivalence on either side of the shell ``wall,'' namely

\begin{align}
    \mu_{i}^2
    = 1 - \left( \frac{n_{i-1}}{n_{i}} \right)^2 \left( 1 - \mu_{i-1}^2 \right),
\end{align}
where n is the index of refraction of O2 and $\theta$ is the angle to the normal of the shell boundary.
Substitution into Equation \ref{eqn:ThetaLi} yields a correction factor onto $\mu_i$.
If the density is low ($\rho \leq 1$ g/cm\textsuperscript{-3}), it can be expected that the refractive index $n$ is near unity \citep{liu2008relationship, phillips1920relation}.
From the Lorenz-Lorentz relation \citep{buckingham1974density}, the refraction index can be expressed relative to the previous refractive index as

\begin{align} \label{eqn:refraction}
    n_{i}
    = \frac{ \eta_{i} }{ \eta_{i-1} } \left( n_{i-1} - 1 \right) + 1,
\end{align}
where $\eta$ is the number density and $n$ is the refractive index.
Using the same low-density assumptions, it is known through the Cauchy formula \citep{cauchy1830, born2013principles} that the dispersion formula for the refractive index as a function of wavelength can be expressed as
\begin{align} \label{eqn:dispersion}
    n_0
    &= A \left( 1 + \frac{B}{\lambda^2} \right) + 1,
\end{align}
where constants A and B are tabulated from observations (see \ref{sec:appendixsimplifyzenith}).

A final complication related to refraction correction arises when we consider the initial zenith angle compared with the ``perceived'' zenith angle.
As shown in Figure \ref{fig:diagram_trig}, the refracted path as perceived by the observer has a given angle $\theta_0$ that is slightly higher than the actual zenith angle.
As such, we must correct the given initial zenith angle such that the curved refraction path aligns with the location of the Sun.
Since the correction for refraction seems to be small for our applications in this paper, it is left for future work.
In the next section, we will put all of our functions together and remove the recursive relationships.
\subsection{Assembling the Model}\label{sec:assemblingthemodel}

In order to avoid computationally expensive functions, we simplify our summations and nested expressions.
Since we are using a uniform atmospheric density model, apply a uniform $\Delta r$.
Expanding the nested functions is rather trivial and results in a much cleaner and faster computation (see \ref{sec:appendixsimplifyzenith} for complete derivation).

\begin{align}
    \Delta L_i
    &= \sqrt{ \left[ r_0 + \left( i + 1 \right) \Delta r_{0} \right]^{2} - \left( r_{0} + i \Delta r_{0} \right)^{2} \left( 1 - \mu_i^2 \right) } - \left( r_{0} + i \Delta r_{0} \right) \mu_i,  \label{eqn:finalequation} \\
    \mu_i^2
    &= 1 - \left( \frac{ r_0 + \Delta r_{0} }{ r_0 + \left( i + 1 \right) \Delta r_{0} } \right)^2 \left( \frac{ n_{0} }{ n_{i} } \right)^2 \left( 1 - \mu_{0}^2 \right), \label{eqn:finalequationzenith} \\
    n_{i}
    &= \frac{ \eta_{i} }{ \eta_{0} } \left( n_{0} - 1 \right) + 1.  \label{eqn:finalequationrefraction}
\end{align}

This equation holds true for any $i \geq 0$.
To avoid confusion with the symbols, recall again that the refractive index is $n$ and the number density is $\eta$.
Among the initial conditions needed are the initial altitude $r_0$ and the cosine of the zenith angle $\mu_0$, which are given based on the location and time of day.
As mentioned in Section \ref{sec:refraction}, the refractive index $n_0$ should be a function of wavelength.
This correction is ignored for now, and in fact refraction, in general, does not contribute to the results presented in Section \ref{sec:results} and can be set to a constant of unity.
In the next sections, we discuss the models for number density $\eta$ and the absorption cross-section $\sigma$.
\subsection{Atmospheric Properties}\label{sec:atmosphericproperties}

In order to compute the expected absorption from the atmosphere, we need a model of the density of the material causing the absorption.
We could begin by setting up the number density to follow the general exponential trend.
However, a complex data-driven model has already been developed by the Naval Research Laboratory: the NRLMSISE-00 Model \citep{NRLMSISE,Bender2020PyNRLMSISE00}.
This model provides users with the temperature and density for Hydrogen, Helium, Oxygen, O\textsubscript{2}, Nitrogen, N\textsubscript{2}, and Argon for specified dates and locations as a function of altitude.

\begin{figure}
    \centering
    \includegraphics[width=0.8\linewidth]{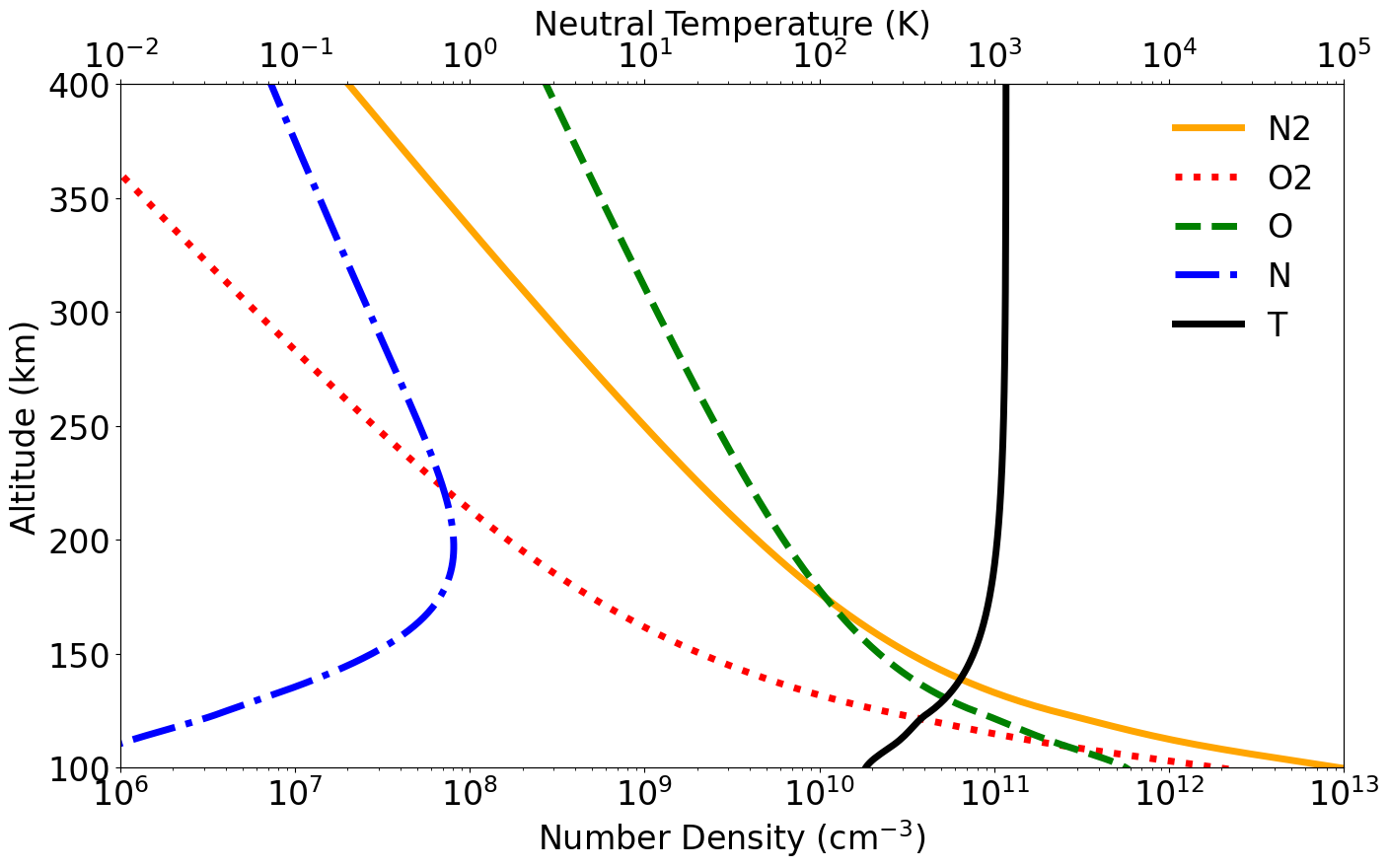}
    \caption{Densities of H, He, O, O\textsubscript{2}, N, N\textsubscript{2}, and Ar.
    These data show good agreement with \cite{meier1991ultraviolet}, Figure 30.
    }
    \label{fig:Density}
\end{figure}

The data plotted in Figure \ref{fig:Density} is retrieved at the time of year and for the specified launch location from \cite{meier1991ultraviolet}.
We can adjust the time of day as well to see the effect this has on the final model.
Similarly, the zenith angle can be found in online tools such as the Solar Position Calculator provided by NOAA \citep{NOAA_ESRL_SolPosCalc_Old, NOAA_ESRL_SolPosCalc_New} or the python integrated package ``PySolar'' \citep{brandon_stafford_2018_1461066}.
For this model, at solar noon $\mu \approx 0.7$ at ground level.

\begin{figure}
    \centering
    \includegraphics[width=\linewidth]{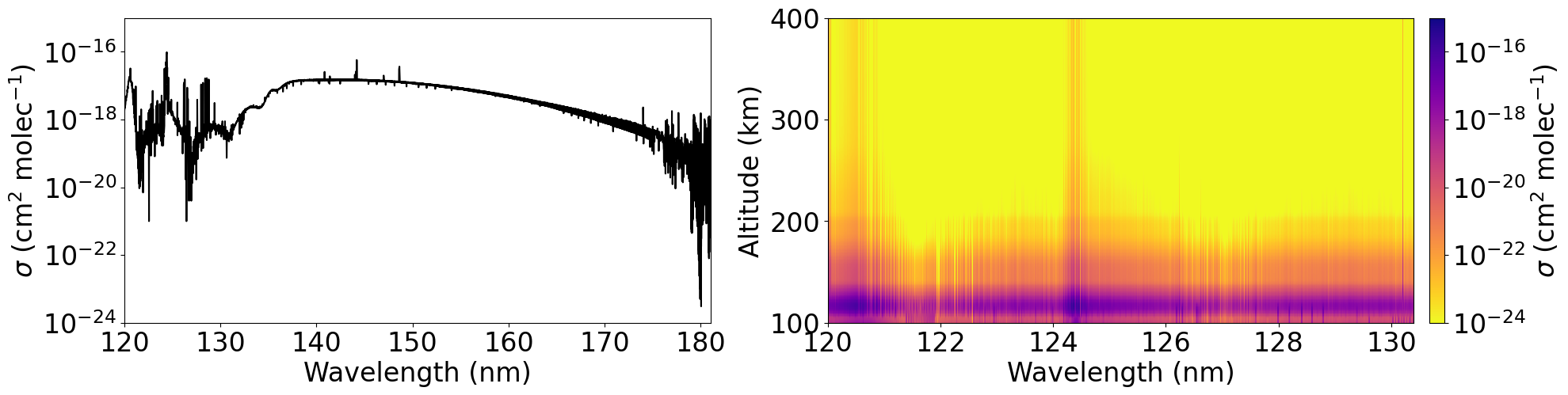}
    \caption{Absorption cross section for O\textsubscript{2} between 120-181 nm (left) and varying with Altitude between 120-130.4 nm.
    These data show agreement with \cite{meier1991ultraviolet}, Figure 27.
    }
    \label{fig:CrossSection}
\end{figure}

The final piece of the optical depth calculation is the absorption cross-section.
We show in Figure \ref{fig:CrossSection} the values for O\textsubscript{2} available from the MPI-Mainz UV-VIS Spectral Database \citep{keller2013mpi, sander2014mpi, MPI_Mainz_Database} for the specified spectral range.
Cross-sections are given in terms of cm\textsuperscript{2} per molecule.
This plot roughly matches the order of magnitude of \cite{meier1991ultraviolet}, Figure 27, as well as a much older paper (\cite{watanabe1953absorption}, Figure 10).

The data set itself is based on laboratory measurements and detailed at various wavelengths and temperatures.
As such, the data are sparse and some interpolation is needed to set up the continuous 2D array shown in Figure \ref{fig:CrossSection} (right).
For this interpolation, we retain the given values while ensuring smooth variation between any missing points.
This is the simplest method that preserves the general trend seen in the data.
As noted later in Section~\ref{sec:inversion}, the lack of complete data is another motivation for detailing this model.
It may be possible to fill in this missing information by looking at raw spectral data taken \textit{in-situ} and determining the absorption cross-section from that measurement.

Note many interesting features here in Figure \ref{fig:CrossSection} (left).
Firstly, the continuum is nicely visible, with a few small peaks near the middle part of the range.
Given a large enough Signal-to-Noise Ratio (SNR), these stand-out peaks could aid in calibration.
The highest value, nearing the -16 order at around 125 nm, will be useful as well.
Finally, there are the Schumann-Runge bands at the far end, beyond 175 nm.
This messy region could provide fiducials for spectral calibration \citep{yoshino1984atlas}.

\begin{figure}
    \centering
    \includegraphics[width=0.8\linewidth]{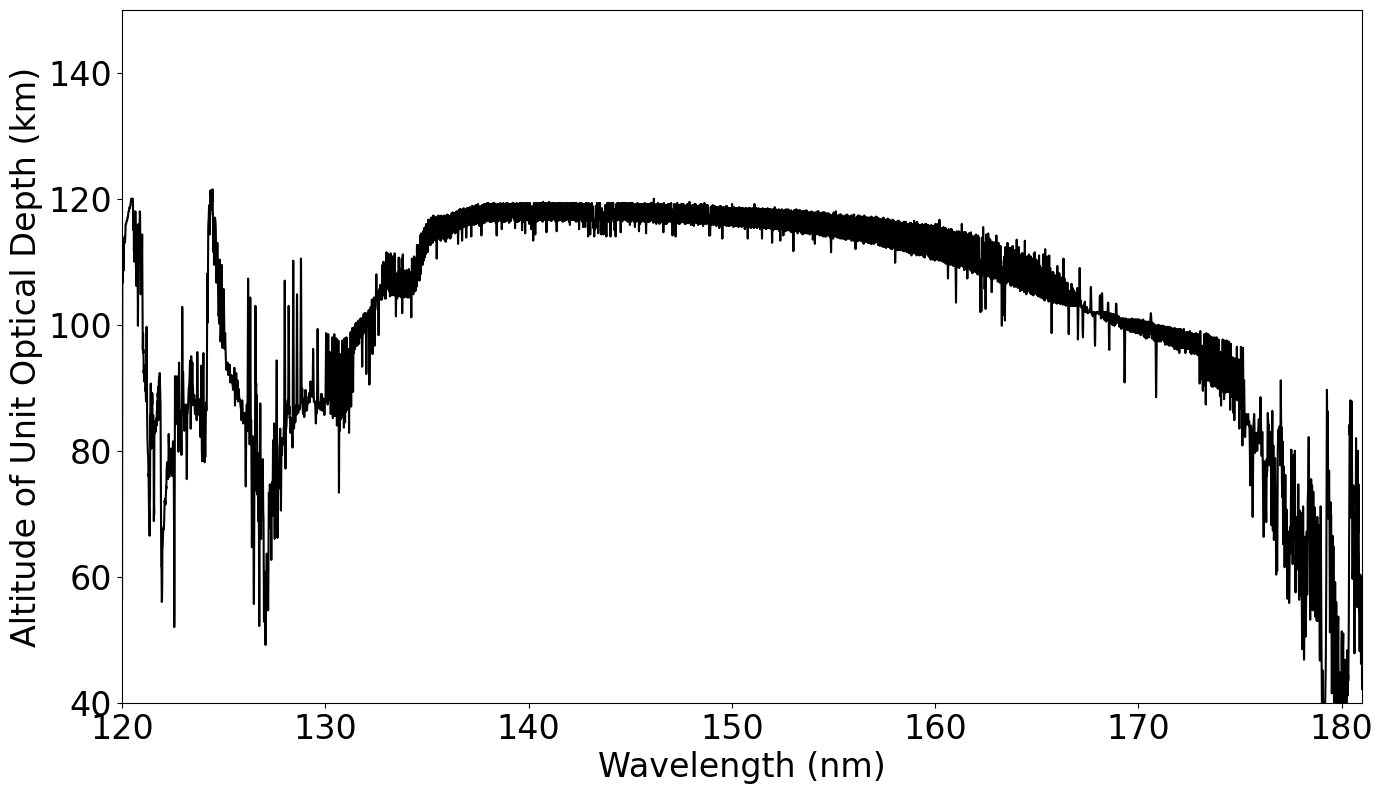}
    \caption{Altitude of Unit Optical Depth as calculated by Equation \ref{eqn:opticaldepth} and \ref{eqn:finalequation}.}
    Shows agreement with \cite{meier1991ultraviolet,lean1987uvirradiance}
    \label{fig:unit_optical_depth}
\end{figure}

Putting the cross-section data together with the number density is the final piece to calculating the optical depth as it varies with height and wavelength $\tau(r, \lambda)$.
A useful representation of this result is the altitude of ``unit optical depth.''
This is the location where $\tau = 1$ in the exponential term of Equation \ref{eqn:opticaldepth}, and is shown as a function of altitude in Figure \ref{fig:unit_optical_depth}.

This result is in good agreement with \cite{meier1991ultraviolet}, which also matched previous results from \cite{lean1987uvirradiance}.
This plot makes sense when compared with the sharp drop in the density of O\textsubscript{2} in Figure \ref{fig:Density}.
Additionally, the right plot in Figure \ref{fig:CrossSection} shows that this altitude has a significant shift in absorption cross-section.
These effects combine to produce a sharp transition at around 80-120 km.
The sharp peaks near the start of the range will certainly provide some calibration points, but the Schumann-Runge bands will most likely be too low for this specific test.
In the next section, we simulate a set of noisy camera images for testing and analysis of this model.
\subsection{Simulated Spectra}\label{sec:simulatedspectra}

After calculating the optical depth, we can simulate an expected CCD signal by combining the transmission value $T$ with the estimated photon flux $I_0$.
In this section, we will explain the method for generating a simulated solar spectrum by using data from the High-Resolution Telescope and Spectrograph (HRTS) as a baseline for solar activity.

The HRTS spectrum is well known and is one of the better databases for the UV range \citep{brekke1993ultraviolet}.
Other sources exist such as AIA and LISARD, which present solar monitoring information for the general public \citep{LISIRD}.
As the name suggests, HRTS has a high spectral resolution (0.005 nm).
This is necessary if we want to use it to predict what FURST will see, which is expected to have a similar or better resolution.

The Flare Irradiance Spectral Model (FISM2) is also used.
Described in \cite{chamberlain2020fism2} and available on \cite{LISIRD}, this model puts together actual measurements alongside an empirical model of solar variability to produce a more complete data set of UV spectral irradiance at 0.1 nm resolution.

\begin{figure}
    \centering
    \includegraphics[width=0.8\linewidth]{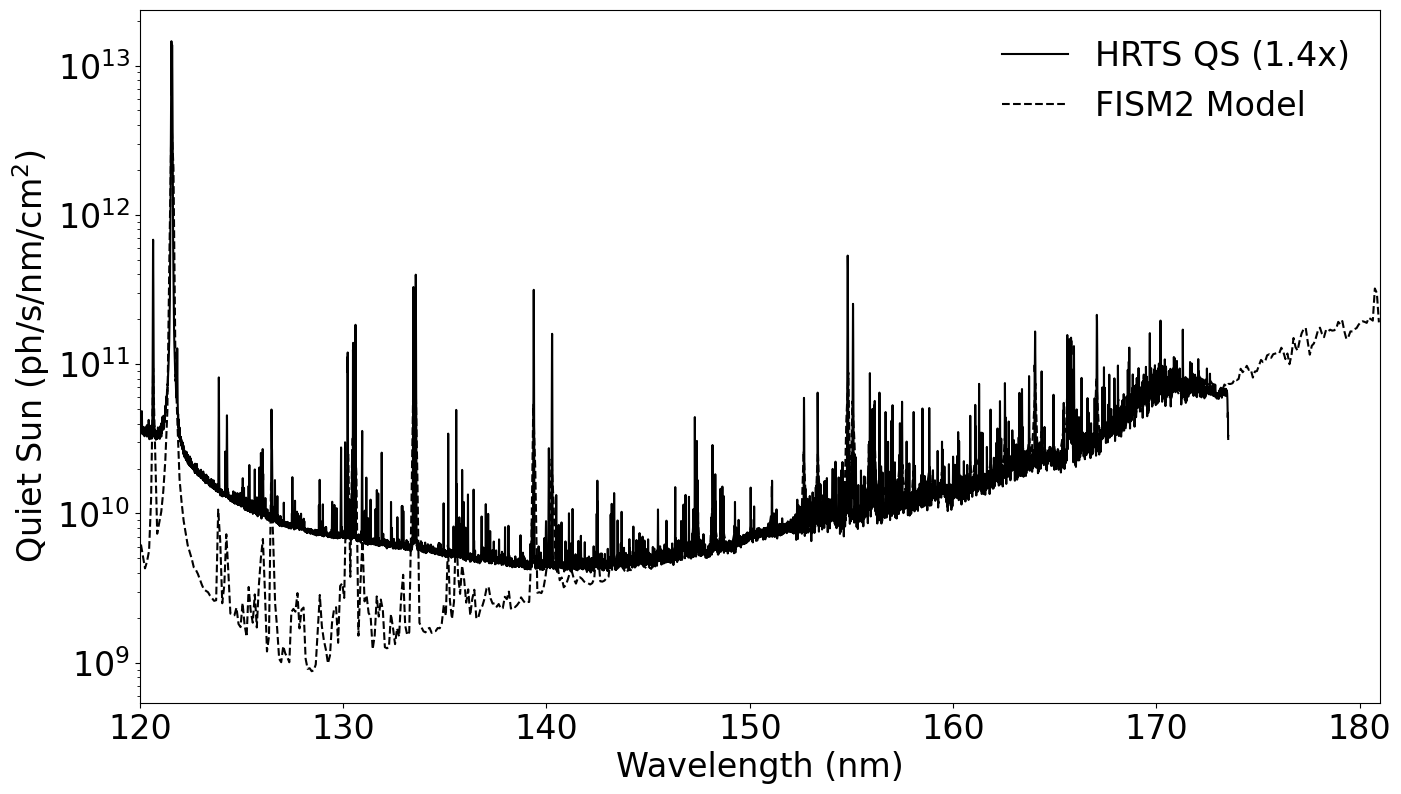}
    \caption{The solid line shows the HRTS spectra for the Quiet Sun (QS) scaled by 1.4x to match the FISM2 model shown by the dashed line.}
    \label{fig:HRTS_RawSpectra}
\end{figure}

The intensity of the HRTS Quiet Sun (QS) signal is shown in Figure \ref{fig:HRTS_RawSpectra}.
Since the noise floor for HRTS is high, we also use the FISM2 model as a reference from the same date (March 1980).
The HRTS spectra shown here are originally given in units of erg/s/sr/\AA/cm$^{2}$.
We convert to ph/s/nm/cm$^2$ by using the size of the full disk ($9.352 \cdot 10^{-3}$ sr), but an additional factor of $\approx$ 1.4x is used in Figure \ref{fig:HRTS_RawSpectra} to scale the intensity to match the FISM2 model.

This difference is relatively small but could be attributed to the scaling of HRTS to the full disk.
Additionally, the QS spectra do not include additional irradiance from Explosive Events (EEs), Active Regions (ARs), etc.
Even so, we prefer to use the HRTS spectra over the FISM2 model in this application because it is given at a much higher spectral resolution.
The best way to obtain a full-disk high-resolution spectrum would be to measure it directly, which is one of the primary motivations for developing an instrument like FURST.

To simulate the spectral signals we need to fold both measurements into the instrument functions for FURST.
This conversion from the raw-data values to Data Numbers (DNs) is done by having knowledge of the instrument, cameras, filters, and other effects such as noise, and can be found in the \ref{sec:appendixtable}.
Of note, there will be a spectral plate scale of $\approx 0.005$ nm per pixel.
This defines the 1-dimensional linear wavelength-to-pixel mapping function used in this paper to demonstrate the model's use for calibration.

Additionally, specifying an exposure time requires us to implement the expected flight profile, since a single exposure occurs at various altitudes.
For this setup, we have a 10-second exposure window at the apogee near 254 km, where there is effectively no absorption, and another at the bottom end of the flight between 115-132 km, where there is much more absorption.

\begin{figure}
    \centering
    \includegraphics[width=0.8\linewidth]{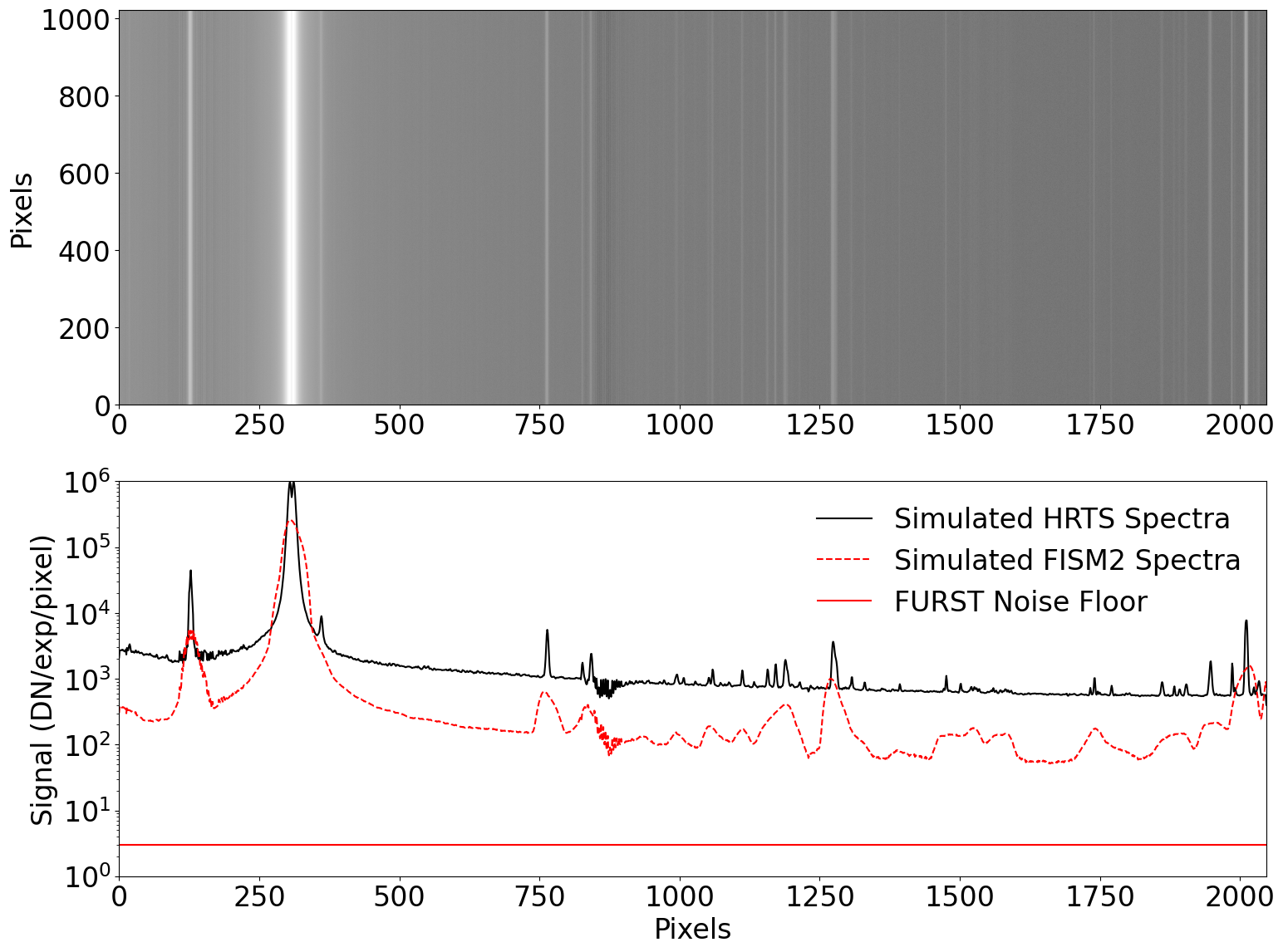}
    \caption{Top: A simulated 2D noisy image from a 10-second exposure between 115-132 km. Bottom: The median signal level across each row (upper/solid black line) with the estimated solar continuum (middle/red dashed line) and the estimated noise floor (lower/solid red line) for the FURST instrument.}
    \label{fig:2Dimage}
\end{figure}

With these conversions in place, a noisy image is generated for a FURST camera signal.
The top image in Figure \ref{fig:2Dimage} is the 2D simulated signal for the window between 115-132 km.
The noise can be seen here, but the spectral features are clear in the bottom plot.
The top solid line is the FURST signal averaged across all rows from the simulated 2D image.
The red dashed line beneath it contains the FISM2 simulated signal as described in Figure \ref{fig:HRTS_RawSpectra}.

An important comparison to this available data is the estimated FURST noise floor.
To calculate this value, we take the various sources of noise from Table \ref{tab:units} in \ref{sec:appendixtable} and work out the highest resulting output signal from an input signal of zero photons.
This works out to an output of approximately 2.5 DN shown as the bottom solid red line in Figure \ref{fig:2Dimage}.

On average, the Standard Error of the Mean (SEM) for these signals is much less than 1\% so it is not displayed here.
It is also encouraging to see that both signals are many orders of magnitude higher than the estimated noise floor for FURST.
This helps to point out the clear improvements being made with the instrumental capabilities FURST will provide.

With the appropriate conversions and noise accounted for, we now have a method for computing atmospheric absorption at the altitudes along a sounding rocket flight profile.
This will allow us to see if the instruments have enough SNR and spectral resolution to be useful for calibration or inversion of atmospheric properties.

\subsection{Calibration}\label{sec:calibration}

Before a spectral image from an instrument can be properly analyzed for solar physics, it must first be calibrated.
There are two different types of calibration of interest to us here: radiometric and spectral.
In this context, radiometric calibration refers to determining the conversion of the signal at each pixel from the computer signal in terms of data number (DN) back into physical units (W cm\textsuperscript{-2} s\textsuperscript{-1} sr\textsuperscript{-1} \AA\textsuperscript{-1}).
Likewise, spectral calibration involves determining the fitting parameters for recasting the pixel numbers in terms of their known wavelength values. 
One common parameter is the spectral plate scale (in nm per pixel), although typically mapping functions are 2D nonlinear functions.

\begin{figure}
    \centering
    \includegraphics[width=\linewidth]{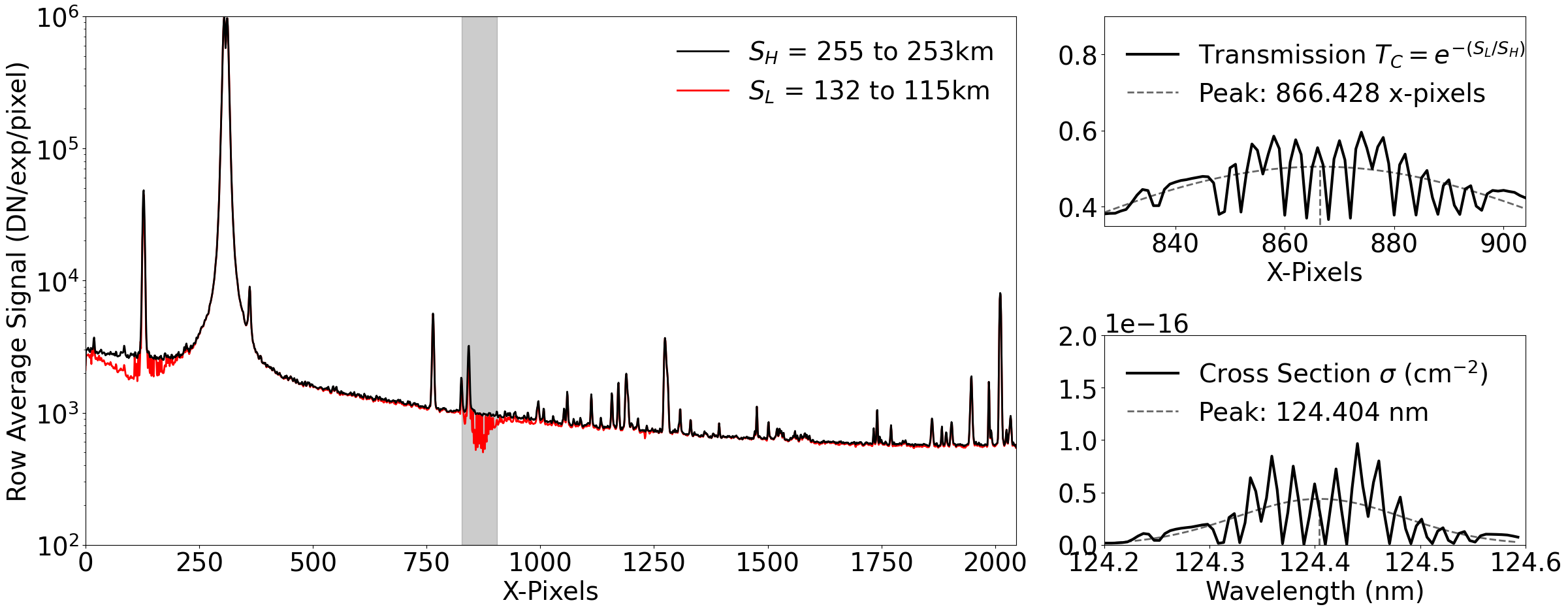}
    \caption{Simulated signal from 10-second exposures, row-averaged. The left plot shows the full wavelength range for one channel.
    The top right is the transmission factor of a zoomed-in region, calculated as shown from the ratio between the simulated data at the higher altitude ($S_H$) and the lower altitude ($S_L$). The lower right plot compares this segment with a known peak in absorption cross-section.}
    \label{fig:calibration}
\end{figure}

To use the identified absorption features for spectral calibration, we use a spherical shells model to properly account for the levels of expected noise and test whether or not we have sufficient resolution to be able to fit these absorption features.
In Figure \ref{fig:calibration}, we show an example of this.
Here we see the simulated signals from a higher altitude $S_H$ and a lower altitude $S_L$.
In the right plots, we compare the ratio of those signals with the absorption cross-section plot, using the corresponding altitude.

A simple Gaussian is used to map out corresponding pixel and wavelength pairs.
From just two of these pairs, we recalculate the spectral plate scale as to within $\approx$0.5\% error.
More sophisticated wavelength mapping programs are to be used in practice to develop our calibration beyond just solving for the linear parameter.
This is a promising result that shows we indeed have sufficient SNR for a viable \textit{in-situ} calibration, as well as that these features are ubiquitous across the spectrum.
\subsection{Inversion}\label{sec:inversion}

Looking back to the models used to generate the spectral images in the previous sections, the NRL density model is well-trusted but the absorption cross-section database is sparse.
This motivated an investigation into a specific application for sounding rockets.
It may be possible to invert the atmospheric absorption from spectral images taken during a flight.

From Equation \ref{eqn:opticaldepth}, the density and absorption cross-section is folded into the integral that yields optical depth.
If spectral images are taken over a relatively small range of altitudes, temperature variations will be small and the cross-section $\sigma$ can be considered independent of height.
Thus, the term can be separated following
\begin{align} \label{eqn:inversion}
    T_{i,j}
    &= e^{- \sigma_{j} \sum_{i'=i}^{\infty} \eta_{i',j} \Delta L_{i}}
    \implies
    \frac{ - \ln{T_{i,j}} }{ \sum_{i'=i}^{\infty} \eta_{i',j} \Delta L_{i} },
    = \sigma_{j},
\end{align}
where the transmission factor $T$ must be found at altitude $i$ and wavelength $j$.
If a spectral image is taken near apogee then absorption is effectively negligible.
Taking the ratio of an observation from a lower height with one taken at the maximum height will give a proxy of $T$ at that lower altitude $i$.
Then we can apply Equation \ref{eqn:inversion} to invert the absorption cross-section using our model.

We can test this with the simulated results from FURST described in Section \ref{sec:calibration} and shown in Figure \ref{fig:calibration} (left).
We take the simulated noisy spectral images taken at the corresponding highest ($\approx$ 254 km) and lowest ($\approx$ 115-132 km) altitudes.
Aside from the noise, the main difference between these two spectral images can be attributed to the atmospheric absorption effect.
We then divide the lower image from the upper images, as in Equation \ref{eqn:inversion}, and recover a ``measured'' optical depth.

The results of this test are shown in Section \ref{sec:results}, Figure \ref{fig:comparingcrosssections}.
As an alternative test, we could place more trust in the cross-section data and attempt to solve for a density profile.
The results of this experiment are shown in \ref{sec:appendixinversiondensity}.
\section{Results} \label{sec:results}

\begin{figure}
    \centering
    \includegraphics[width=0.8\linewidth]{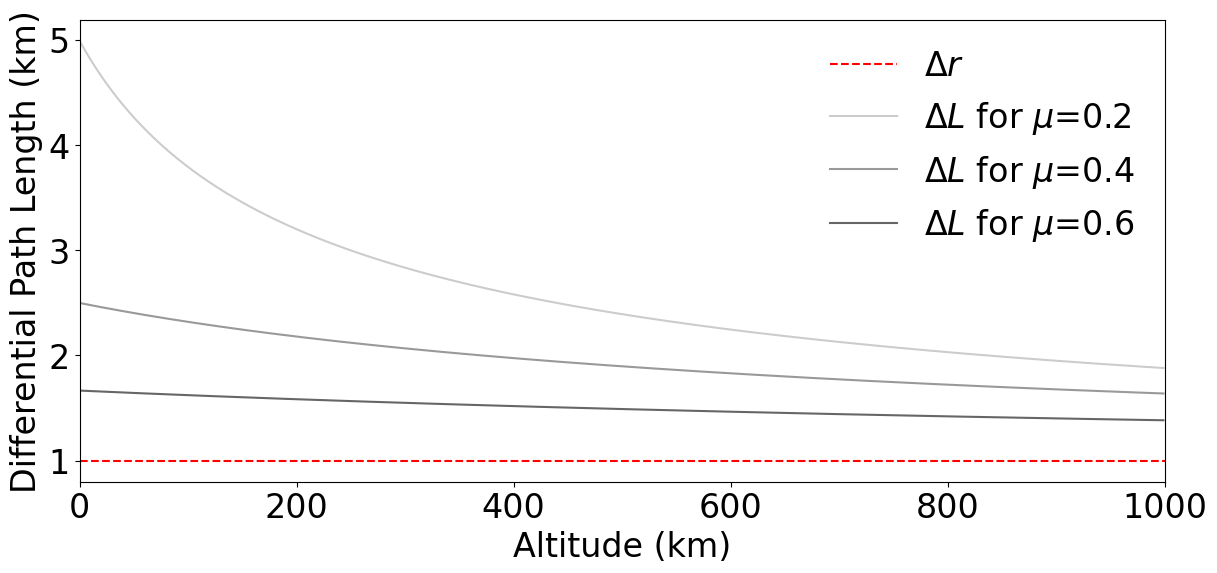}
    \caption{Comparison of the Differential Path Length $\Delta L$ at different initial zenith angles $\mu$ in shades of gray.
    The bottom red dashed line indicates the altitude resolution $\Delta r$ of 1 km in this example.
    The simple model would calculate $\Delta L$ as $\Delta r / \mu$, which corresponds to the initial point on the left, at 0 km.}
    \label{fig:comparing_models}
\end{figure}

We first show a comparison of the differential path length as calculated by the ``simple'' model (where $\Delta L = \Delta r / \mu$) with the result from the spherical-shells model (where $\Delta L$ is calculated as in Equation \ref{eqn:finalequation}).
An example is shown in Figure \ref{fig:comparing_models}, generated at varying initial zenith angles ($\mu_0 =$ 0.2, 0.4, and 0.6) to highlight the effect.
We can see the difference between the use of uniform and non-uniform differential path lengths from the simple model and the shells model, respectively.

The simple model would estimate a uniform differential path length as $\Delta r / \mu$, which corresponds to the initial points (at 0 km) in Figure \ref{fig:comparing_models}.
However, the spherical shells model allows for this to be a non-uniform value.
At higher shells, the angle between the shell and the path approaches normal ($\mu \longrightarrow 1$), forcing the value of the differential path length to approach the shell spacing ($\Delta L \longrightarrow \Delta r$) (bottom/red dashed line).
To help understand the effect of this, we can compare the sum of the $\Delta L$ values between each model.
When $\mu \approx 0.2$, the shells model puts this value at about half of the simple model.
As we approach a zenith angle greater than $0.6$ the difference drops to less than 10\%.
This model is still in-line with the previous work by \cite{meier1991ultraviolet} since the time of day provided a high zenith angle.
These results, however, predict slightly less absorption overall for our flight path than simpler models.

Next, as can be seen in Figures \ref{fig:2Dimage} and \ref{fig:calibration}, once the signal is transmitted through the instrument we have enough SNR to tell where the absorption lines are.
These fiducials provide calibration points to verify the pixel-to-wavelength mapping of the instrument to within 0.5\%, even though we only considered 2 lines in this basic example.
Furthermore, this calibration only considered one wavelength channel for FURST.
By coupling more fiducials and a broader range we hope to add significantly to the wavelength calibration of the data through this \textit{in-situ} method.

Finally, we solve for the absorption cross-section $\sigma$ required to produce the spectra seen by an instrument like FURST.
As described in Section \ref{sec:inversion}, this inversion required taking the ratio of the spectral image taken at apogee with one from a lower height (see Figure \ref{fig:calibration}) and is formulated in Equation \ref{eqn:inversion}.

\begin{figure}
    \centering
    \includegraphics[width=0.7\linewidth]{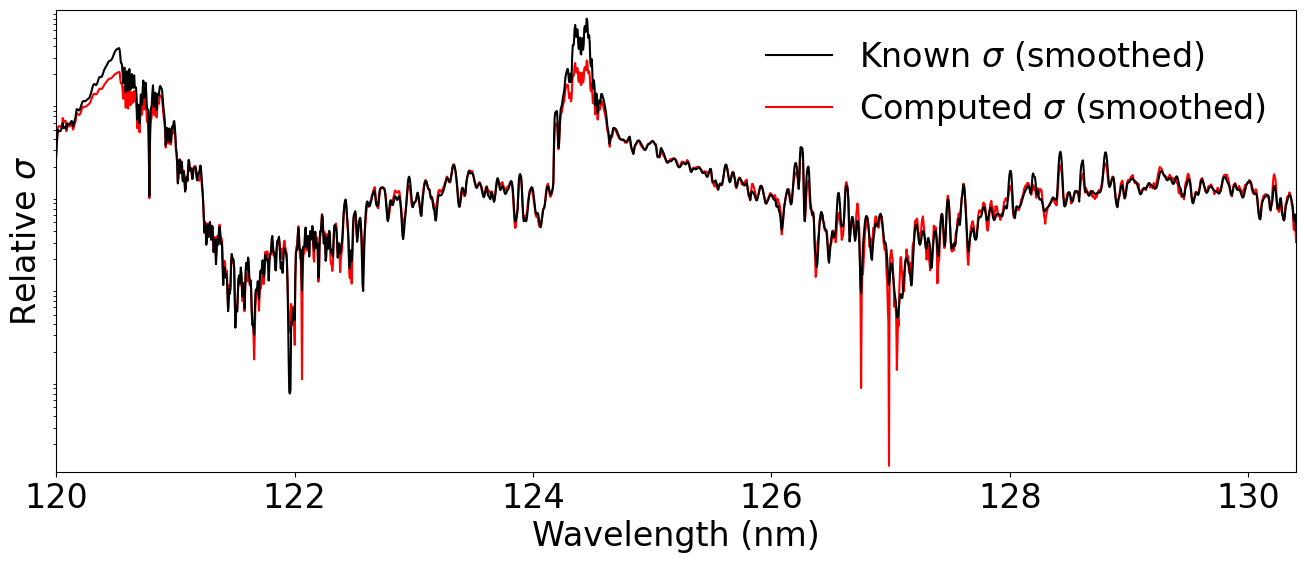}
    \caption{Recalculated absorption cross-section from simulated images (red) compared with the known data (black). Both are smoothed and scaled for easier comparison.}
    \label{fig:comparingcrosssections}
\end{figure}

In Figure \ref{fig:comparingcrosssections}, the black and red lines represent the known and the computed absorption cross-section $\sigma$, smoothed and scaled for ease of comparison.
We clearly reproduce the values quite well, but it is also apparent where noise has made an impact.
There are a few deep lines (around 122 nm and 127 nm) that are a result of this.

Additionally, the absolute values are a bit off.
From our testing, this can be attributed to the temperature change at the altitudes where this simulated signal occurs
There is a sharp change in the temperature, and thus the absorption cross-section values.
To better reproduce the laboratory values, we can employ a more nuanced calculation.
For example, we have utilized a weighted average based on the flight profile.
However, more work must be done to understand what is missing.

Regardless, the primary features of interest are the peak locations rather than their absolute values, and this result shows that they are almost completely recovered.
There is also a tendency for higher values to have less error.
This is because there is less noise associated with the spectral image signal from which it is calculated.
It may not be possible to perform this calculation from instruments with a low SNR.
This further stresses the importance of radiometric calibration, and the added value such an effort brings.

This method shows a broadly applicable technique for the inversion of atmospheric properties from existing data.
It may help significantly to fill gaps in the predicted absorption points in our databases.
We have set up this model to look at molecular oxygen (O$_2$) since this is the primary contributor to UV absorption.
However, other molecules can be explored with this model that may be of greater interest to atmospheric scientists (e.g. Ozone, Methane, Aerosols, etc.).
\section{Discussion} \label{sec:discussion}

This updated model has been shown to accurately reproduce the expected atmospheric absorption predicted by previous models using modern techniques.
Our results match previous reports well while accounting for the curvature of the Earth and its effects on a non-uniform differential path length.
For zenith angles above $\mu > 0.6$, a simple uniform differential path length will change the optical depth by a factor of less than 10\%.
This model is especially suited to applications with a low zenith angle, such as heliophysics tests performed outside of solar noon, at high latitudes, or with a low flight apogee.
With knowledge of the proper absorption effects, this model can assist in extending the launch capabilities for such missions.

This model's biggest benefit for our team, though, is in improving our understanding of the UV absorption effect in FURST data products.
It will also enable us to set limitations or requirements for future launch conditions.
Missions with a desire to measure or avoid UV absorption can benefit from such understanding.

There are of course many improvements that can be made.
Extrapolating the given density data set out above 1000 km, and increasing the resolution, could improve the resolved density profile or better predict the absorption amount.
There also remain other atmospheric effects that we anticipate being useful for the calibration and inversion of the FURST data products.
Hydrogen and Oxygen resonant absorption adds additional absorption features that can be utilized for calibration and cross-section inversions.
A thermally expanded atmosphere due to solar sub-storms, or other such responses to flare events, will have the effect of increasing the altitude of unit optical depth \citep{Jacchia1966diurnal}.

The largest improvement to be made is with the sparseness of the absorption cross-section data.
Around the upper heights of the rocket flight, the temperature does not vary all that much.
However, the NRL model shows significant variation in temperature at the altitudes corresponding to the beginning and end of the shutter open time.
We have attempted to use the available data and have adjusted our code to accept whatever new data becomes available.
Rather than rely on laboratory measurements, we could look into calculating the theoretical spectral width of the O$_2$ absorption features.
Not only should spectral widths change with temperature, but some lines may only appear at certain temperature ranges.
Understanding the physics behind calculating the resonant absorption lines will likely lead to an understanding of how to solve this problem.

Additional work on FURST involves improved radiometric calculations for tracking errors throughout the calibration system and flight instrument and has been discussed in a separate paper \citep{vigil2021design}.
With these high-resolution capabilities, the noise floor of the FURST instrument will be well below the expected solar continuum, making the absorption profile easier to resolve.
Following our spherical shells model, \textit{in-situ} wavelength calibration and radiometric inversion of the atmospheric absorption cross-section should be possible.

\section{Acknowledgements}

The primary author wishes to thank their advisors, mentors, and colleagues that helped with countless meetings, advice, and support along the way: Amy, Charles, Larry, Gen, Adam, Gary, Michael, Keyvan, Clayton, Juan, all of the staff at UAH for the funding support, his wife Crystal, but most importantly, God (Matthew 6:33-34).
This material is based upon work supported by the NSF EPSCoR RII-Track-1.2a (Non-invasive plasma diagnostics for LTP) Cooperative Agreement OIA-1655280. Any opinions, findings, conclusions, or recommendations expressed in this material are those of the authors and do not necessarily reflect the views of the National Science Foundation.
\bibliography{main.bib}
\appendix \section{Details of Differential Path Length Formula} \label{sec:appendixdiffpathlength} \setcounter{section}{1}

The following shows more details in the derivation of the differential path length calculations.
Refer to Figure \ref{fig:diagram_trig} as needed.
To solve for the differential path length $\Delta L_{i}$, we first employ the Law of Sines.
Firstly, since $\sin \left( \pi - \theta_0 \right) = \sin \theta_0$, then

\begin{align}
    \frac{ \sin \theta_0 }{ r_0 + \Delta r_0 }
    &= \frac{ \sin \left( \theta_{L0} \right) }{ r_0 }
    \implies \theta_{L0}
    = \sin^{-1} \left( \frac{ r_0 }{ r_0 + \Delta r_0 } \sin \theta_0 \right).
\end{align}

Then we use the 180\textdegree\ rule to solve for $\theta = \theta_0 - \theta_{L0}$.
Finally, after some algebra, we use the Law of Sines again to solve for $\Delta L_0$.

\begin{align}
    \frac{ \sin \theta_0 }{ r_0 + \Delta r_0 }
    = \frac{ \sin \theta }{ \Delta L_0 }
    \implies \Delta L_0
    &= \sqrt{ \left( r_0 + \Delta r_0 \right)^{2} - r_{0}^{2} \left( 1 - \mu_0^2 \right) } - r_0 \mu_0,
\end{align}
where we allow the convention of $\mu_0 = \cos \theta_0$.
Extending to each subsequent shell, we can see the pattern for $\Delta L_{i}$ as

\begin{align}
    \Delta L_0
    &= \sqrt{ \left( r_0 + \Delta r_0 \right)^{2} - r_{0}^{2} \left( 1 - \mu_0^2 \right) } - r_0 \mu_0 \\
    \Delta L_1
    &= \sqrt{ \left( r_0 + \Delta r_0 + \Delta r_1 \right)^{2} - \left( r_{0} + \Delta r_0 \right)^{2} \left( 1 - \mu_1^2 \right) } - \left( r_{0} + \Delta r_0 \right) \mu_1 \\
    \Delta L_2
    &= \sqrt{ \left( r_0 + \Delta r_0 + \Delta r_1 + \Delta r_2 \right)^{2} - \left( r_{0} + \Delta r_0 + \Delta r_1 \right)^{2} \left( 1 - \mu_2^2 \right) } - \left( r_{0} + \Delta r_0 + \Delta r_1 \right) \mu_2 \\
    \vdots \nonumber \\
    \Delta L_i
    &= \sqrt{ \left( r_0 + \sum_{i'=0}^i \Delta r_{i'} \right)^{2} - \left( r_{0} + \sum_{i'=0}^{i-1} \Delta r_{i'} \right)^{2} \left( 1 - \mu_i^2 \right) } - \left( r_{0} + \sum_{i'=0}^{i-1} \Delta r_{i'} \right) \mu_i, \\
    \mu_i
    &= \sqrt{1 - \left( \frac{ r_0 + \sum_{i'=0}^{i-1} \Delta r_{i'} }{ r_0 + \sum_{i'=0}^{i} \Delta r_{i'} } \right)^2 \left( 1 - \mu_{i-1}^2 \right) } ,\ i > 0, \label{eqn:cosmuappendix} \\
    \mu_0
    &= \cos \theta_0.
\end{align}

From Snell's law of refraction, there exists an equivalence on either side of the shell ``wall,'' namely

\begin{align}
    n_{i} \sin \theta_{i}
    &= n_{i-1} \sin \theta_{i-1}
    \implies \mu_{i}^2
    = 1 - \left( \frac{n_{i-1}}{n_{i}} \right)^2 \left( 1 - \mu_{i-1}^2 \right),
\end{align}
where n is the index of refraction of O$_2$ and $\theta$ is the angle to the normal of the shell boundary.
Upon substitution into Equation \ref{eqn:cosmuappendix}, we end up with a correction factor onto $\mu_i$ given by the ratio of the index of refraction at each subsequent shell (${n_{i-1}}/{n_{i}}$).
The relationship between density and refractive index can be expressed linearly if $\rho \leq 1$ g/cm\textsuperscript{-3} (see \cite{liu2008relationship, phillips1920relation}, and others).
For O$_2$, this corresponds to a number density $\eta_{O2} \leq 10^{22}$ atoms/cm\textsuperscript{3}).
If the density is low, it can be expected that the refractive index $n \rightarrow 1$.
From the Lorenz-Lorentz relation \cite{buckingham1974density} we can calculate this as

\begin{align} \label{eqn:refractionappendix}
    \frac{n^2 - 1}{n^2 + 2} \rho^{-1}
    &= \frac{N_A \alpha}{3 \epsilon_0}
    \quad
    \overset{n \to 1}{\Longrightarrow}
    \quad
    n
    = \frac{ N \alpha_0 }{ 2 \epsilon_0 } \rho + 1.
    \implies n_{i}
    = \frac{ \eta_{i} }{ \eta_{i-1} } \left( n_{i-1} - 1 \right) + 1.
\end{align}

In Equation \ref{eqn:refractionappendix}, $N_A$ and $\epsilon_0$ are constants.
The polarizability $\alpha$ is highly dependant upon density and wavelength.
However, under linearity approximation at low density, the dependency can be expressed as a function of number density $\eta$.
Taking the ratio of refractive indices of subsequent shells removes those constants, and only a single value for the refractive index at the instrument starting location $n_0$ is needed.

Using those same low-density assumptions, it is known through the Cauchy formula \citep{cauchy1830} and \cite{born2013principles} that the dispersion formula for the refractive index as a function of wavelength can be expressed as
\begin{align} \label{eqn:dispersionappendix}
    n_0
    &= A \left( 1 + \frac{B}{\lambda^2} \right) + 1,
\end{align}
where constants A and B are tabulated from laboratory measurements to be $2.663\cdot10^{-4}$ and $5.07\cdot10^{3}$ nm\textsuperscript{2} respectively.
A more precise dispersion relation can be formulated using polarizability functions, but this is not necessary for a first-order solution.
Since we have shown that the refractive index can be expressed as a function of wavelength, this implies that the differential path length is as well.

\section{Simplifying Functions} \label{sec:appendixsimplifyzenith}

There is a fair amount of simplification that occurs when we see how the nested functions begin to cancel out terms.
For the zenith angle, the recursive functions terminate at $i=0$ as follows.
\begin{align}
    \mu_i^2
    &= 1 - \left( \frac{ r_0 + \sum_{i'=0}^{i-1} \Delta r_{i'} }{ r_0 + \sum_{i'=0}^{i} \Delta r_{i'} } \right)^2 \left( \frac{n_{i-1}}{n_{i}} \right)^2 \left( 1 - \mu_{i-1}^2 \right) \\
    &= 1 - \left( \frac{ r_0 + \sum_{i'=0}^{i-1} \Delta r_{i'} }{ r_0 + \sum_{i'=0}^{i} \Delta r_{i'} } \right)^2 \left( \frac{n_{i-1}}{n_{i}} \right)^2 \left( \frac{ r_0 + \sum_{i'=0}^{i-2} \Delta r_{i'} }{ r_0 + \sum_{i'=0}^{i-1} \Delta r_{i'} } \right)^2 \left( \frac{n_{i-2}}{n_{i-1}} \right)^2 \left( 1 - \mu_{i-2}^2 \right) \\
    &= 1 - \left( \frac{ r_0 + \sum_{i'=0}^{i-1} \Delta r_{i'} }{ r_0 + \sum_{i'=0}^{i} \Delta r_{i'} } \right)^2 \left( \frac{ r_0 + \sum_{i'=0}^{i-2} \Delta r_{i'} }{ r_0 + \sum_{i'=0}^{i-1} \Delta r_{i'} } \right)^2 \dots \nonumber \\
    & \quad\quad\quad \left( \frac{ r_0 + \Delta r_{0} }{ r_0 + \Delta r_{0} + \Delta r_{1} } \right)^2 \left( \frac{n_{i-1}}{n_{i}} \right)^2 \left( \frac{n_{i-2}}{n_{i-1}} \right)^2 \dots \nonumber  \\
    & \quad\quad\quad \left( \frac{n_{0}}{n_{1}} \right)^2 \left( 1 - \mu_{0}^2 \right) \\
    &= 1 - \left( \frac{ r_0 + \Delta r_{0} }{ r_0 + \sum_{i'=i}^{i} \Delta r_{i'} } \right)^2 \left( \frac{n_{0}}{n_{i}} \right)^2 \left( 1 - \mu_{0}^2 \right).
\end{align}

Likewise, the formula for refractive index reduces to Equations \ref{eqn:finalequationrefraction}.
To obtain the final expression for $\Delta L$, we invoke the uniform shell-spacing $\Delta r_i = \Delta r$.
Thus, summations $\sum_{i'=0}^{i} \Delta r_i$ become $(i+1) \Delta r$.
This returns Equations \ref{eqn:finalequation} and \ref{eqn:finalequationzenith}.

\section{Conversion of Signal to Data Numbers} \label{sec:appendixtable}

A spectral signal is often given in units of ergs/s/\AA/sr/cm\textsuperscript{2}.
Table \ref{tab:units} shows the conversion process to units Data Numbers per exposure per pixel.
Firstly, we convert the energy from ergs to photons using the photon energy.
Then we have to account for the spectral plate scale for each wavelength channel in FURST, which is $\approx 10.4$ nm across a width of 2048 pixels.
Since we are observing the full disk of the sun ($\approx$32 arc-minutes), then its size in our field of view will be roughly constant at the value shown.

The signal is then attenuated by the instrument optics and the physical size of the CCD.
Finally, the photon count rate is spread across the height of the CCD, which if the optics are aligned properly should match the number of rows of pixels.
In folding the photon signal into the instrument response function, we take into account the charge spreading of the signal across adjacent pixels.
This width $\sigma$ is given from the FURST design and is around a pixel in size depending on the wavelength.
We compute the line-spread function using a standard Gaussian, $I_{0} \cdot \exp( -(\lambda - \lambda_0) / 2 \sigma^{2})$, where $\lambda_0$ is taken as the HRTS signal mapped onto the defined $\lambda$ space.

The camera sends this signal along the readout channels after a set exposure time, which for this testing we have set to 10 seconds.
In order to count up the total number of photons during each exposure, we use the flight profile to fold in the hanging transmission factor, as calculated by Equation 1 and shown in the Results in Figure \ref{fig:unit_optical_depth}.
In either case, it is at this point that the signal is saved as an integer number of photons per exposure per pixel.

\begin{table}[H]
    \centering
    \begin{tabular}{ll|ll}
        Variable               & Source                                        & Values                                   & Units \\ \hline \hline
        HRTS Signal            & Raw Data                                      & $\approx 414^{+11912}_{-307}$            & erg/s/\AA/sr/cm$^{2}$ \\ \hline
        Photon Energy          & E$_{ph}$ = h $\cdot$ c $\cdot$ $\lambda^{-1}$ & $\approx 1.32 \cdot 10^{-11}$            & erg/ph \\
        Spectral Plate Scale   & Optical Setup                                 & $5.078 \cdot 10^{-3}$                    & \AA/x-pixel \\
        Solar Disk Area        & Observation                                   & $9.352 \cdot 10^{-3}$                    & sr \\
        Effective Area         & GA $\cdot$ Refl. $\cdot$ Trans. $\cdot$ QE    & $\approx 4.176 \cdot 10^{-5}$            & cm$^{2}$ \\
        Number of Rows         & Variable                                      & 1024                                     & y-pixel \\
        Exposure Time          & Variable                                      & 10                                       & s/exp \\
        Atmospheric Absorption & Model                                         & 0.0-1.0                                  & unit-less \\ \hline
        Photon Signal          & Converted                                     & $\approx 2070^{+91130}_{-1740}$          & ph/exp/pixel \\ \hline
        Photon Noise           & Poisson                                       & $\approx \pm 110.5 \ (3\sigma)$          & ph/pixel \\
        Electrons per Photon   & E$_{\text{ph}}$/E$_{\text{eh}}$               & $\approx 2.26$                           & e$^{-}$/ph \\
        Electron Noise         & Readout, Fano, Shot, etc.                     & $\approx \pm 25 \ (3\sigma)$             & e$^{-}$/pixel \\
        Gain                   & Variable                                      & 1.0                                      & DN/e$^{-}$/pixel \\
        Gain Bias              & Variable                                      & 0 (will be 3000)                         & DN/pixel \\ \hline
        Camera Signal          & Converted                                     & $\approx 5645^{+258129}_{-4811}$         & DN/exp/pixel
    \end{tabular}
    \caption{A summary of the unit conversions and sources of noise in the signal. Beyond the Photon Signal step, the values are interpreted as integers by the camera and readout systems.}
    \label{tab:units}
\end{table}

Since there can be inherent variability in the count rate, we add Poisson noise.
Likewise, the CCD converts the incoming photons to electrons based on the electron hole-pair Energy in Silicon (E\textsubscript{eh} $= 5.847944 \cdot 10^{-12}$ erg/e\textsuperscript{-}).
The CCD reads out the total electron count from each pixel in terms of voltage, with readout noise, gain, and gain bias.
A separate calibration code interprets the voltage signal to the computer system as a Data Number based on this gain and gain-bias value.
It is not necessary to include a gain-bias value for this testing, since this value doesn't fluctuate and is only present to prevent negative values in the computer processing system.

\section{Inversion of Density Model} \label{sec:appendixinversiondensity}

If we have enough confidence in the absorption cross-section, we can in practice recalculate the density model instead.
Starting as before, we can take the ratio of the apogee image with any lower-altitude image as a proxy of the transmission factor.
We then use a forward model to invert the problem and produce a potential solution for the atmospheric density profile.
To do this, we set up an optimization routine to vary the parameters of an assumed functional form of the density profile.

This profile is carried through the spherical-shells model and compared with the simulated ``measured'' optical depth.
The profile used here was formulated by observation of the NRL model to be a combination of a linear and parabolic line.
Thus, using a conventional approach often used in setting up a broken power law (insert citation), we define the functional form of the density to roughly follow

\begin{align}
    y
    &= \frac{f_1 - f_2}{\pi} \arctan \left( k \left( x_0 - x \right) \right) + \frac{f_1 + f_2}{2},
\end{align}
where y is the density, x is the altitude, and $f_1 = a(x-x_0)^2 + b(x-x_0)+c$ and $f_2 = b(x-x_0)+c$ are the standard parabolic and linear functions, respectively.
The coefficients a, b, c, and x$_0$ are adjusted by the optimization routine.
The value for k adjusts the level of sharpness around the break-point between the two functions and is simply set to a relatively high number such as 1000.

\begin{figure}
    \centering
    \includegraphics[width=0.6\linewidth]{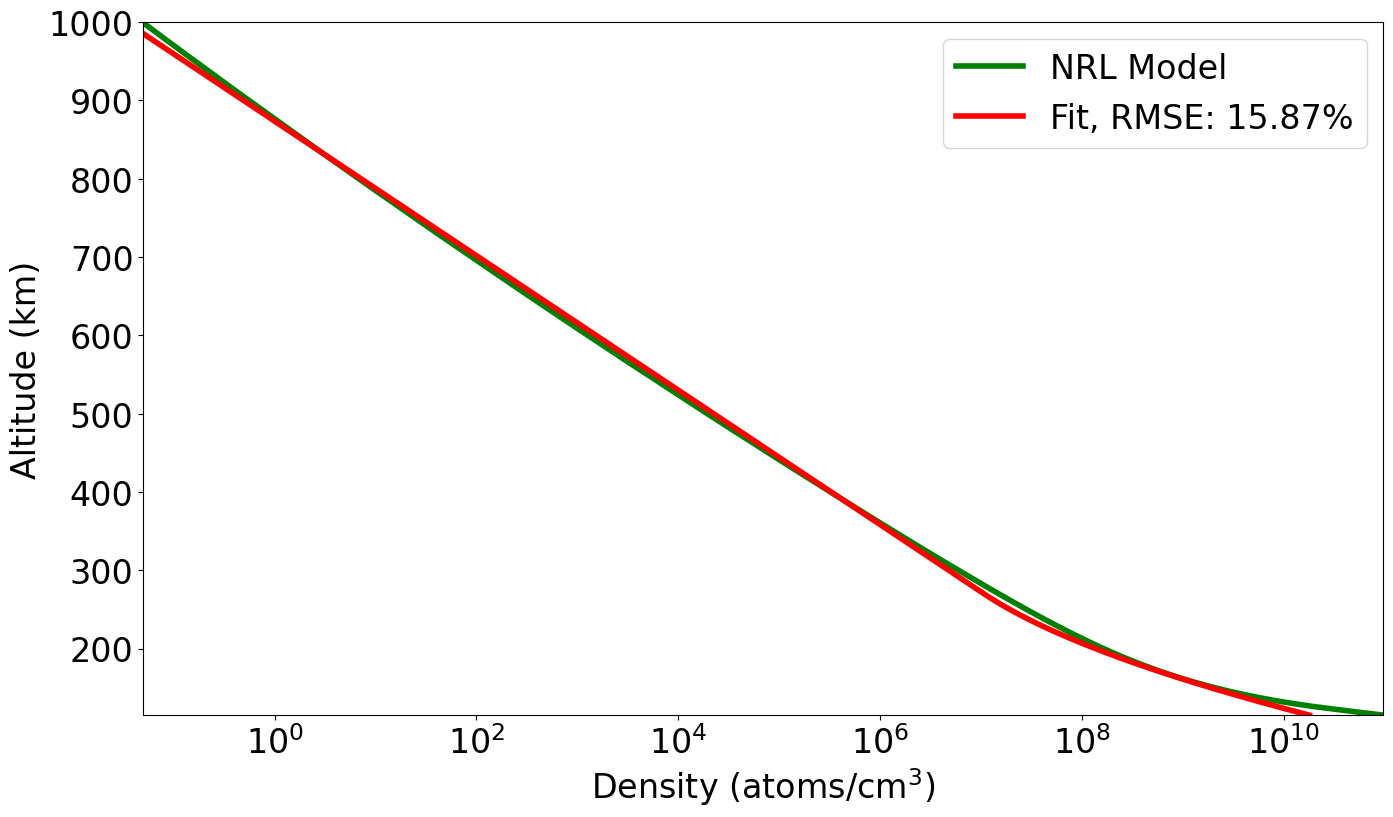}
    \caption{Comparing the density model as used to generate the simulated signal (green line) and the recomputed values through the optimization routine (red line).
    }
    \label{fig:comparing_densities}
\end{figure}

An initial guess for the parameters is given based on a fitting of the NRL model.
We then apply those parameters as the initial guess for computing the density within equation 1.
This result is compared with the signal data, and the optimization routing minimizes the difference between the two by adjusting those initial guesses.
As can be seen in Figure \ref{fig:comparing_densities}, the fit is not perfect but it is close.
From these simulated noisy images we can reproduce the density curve from the NRL model between the observed altitude and 1000km with a Root Mean Squared Error (RMSE) of about 16\%.

We do not consider this routine robust enough, as the initial guess is based on required prior knowledge and complete trust in the sparse absorption cross-section data set.
This step will continue to be improved upon though, and combining observations from various shutters and timings may improve this routine.
Additionally, we may be able to use the solved absorption cross-section from one image to inform this inversion of the data from another image.

\end{document}